\documentclass[a4paper,twoside,10pt,english,numbers=noenddot]{scrbook}

\usepackage[english]{babel}
\makeatletter
\def\month@ngerman{\ifcase\month \or Januar\or Februar\or M\"arz\or April\or Mai\or Juni\or Juli\or August\or September\or Oktober\or November\or Dezember\fi}
\def\month@english{\ifcase\month \or January\or February\or March\or April\or May\or June\or July\or August\or September\or October\or November\or December\fi}
\makeatother
\usepackage{hyphenat} 

\usepackage[T1]{fontenc}
\usepackage{helvet} 
\usepackage{indentfirst}
\usepackage[keeplastbox]{flushend} 

\usepackage[usenames,dvipsnames]{xcolor} 
\usepackage{graphicx} 
\usepackage{wrapfig} 
\usepackage{tikz}
\usepackage{rotating}

\usepackage{ifthen}
\usepackage{forloop}
\usepackage{etoolbox}

\usepackage[fleqn]{amsmath} 
\usepackage{amssymb} 
\usepackage{amsfonts} 
\usepackage{amstext} 
\usepackage{wasysym}
\usepackage{mathrsfs} 
\usepackage{mathtools}
\usepackage{upgreek}
\usepackage{siunitx}
\usepackage{esint} 
\usepackage{cancel} 
\usepackage{empheq}
\allowdisplaybreaks

\usepackage{multicol} 
\usepackage{multirow} 
\usepackage{dcolumn}
\usepackage{tabularx} 
\newcolumntype{L}[1]{>{\raggedright\arraybackslash\hsize=#1\hsize}X}
\newcolumntype{R}[1]{>{\raggedleft\arraybackslash\hsize=#1\hsize}X}
\newcolumntype{C}[1]{>{\centering\arraybackslash\hsize=#1\hsize}X}
\usepackage{dblfloatfix} 

\usepackage{enumitem} 
\setlist{nosep} 

\usepackage[a4paper,portrait]{geometry}
\newlength{\TwoColumnWidth}
\newlength{\OneColumnWidth}
\usepackage[headsepline]{scrlayer-scrpage}

\usepackage[hang,bottom]{footmisc}
\deffootnote{\footnotemargin}{0pt}{%
	\textsuperscript{\thefootnotemark}
}
\usepackage[singlelinecheck=false,font=small,labelfont=bf,format=plain]{caption} 
\usepackage[singlelinecheck=true,font=small,labelfont=bf,format=plain]{subfig}
\setkomafont{sectioning}{\rmfamily\bfseries}
\setkomafont{descriptionlabel}{\rmfamily\bfseries}

\usepackage[subfigure]{tocloft} 
\addto\captionsenglish{

}
\makeatletter
  \renewcommand*{\@pnumwidth}{20pt} 
  \renewcommand*{\@tocrmarg}{30pt plus 5pt minus 0pt} 
  \renewcommand*{\@dotsep}{4} 
\makeatother

\newcounter{chapterappendixcounter}[chapter]

\newcounter{totalpagecounter}
\newcounter{totalfigurecounter}
\newcounter{totaltablecounter}
\newcounter{totalcitecounter}
\newcounter{totalpages}
\newcounter{totalfigures}
\newcounter{totaltables}
\newcounter{totalcites}
\makeatletter
\@input{\jobname.myaux}
\makeatother

\definecolor{white}{rgb}{1,1,1}
\definecolor{black}{rgb}{0,0,0}
\definecolor{red}{rgb}{1,0,0}
\definecolor{green}{rgb}{0,1,0}
\definecolor{blue}{rgb}{0,0,1}
\definecolor{cyan}{rgb}{0,1,1}
\definecolor{magenta}{rgb}{1,0,1}
\definecolor{yellow}{rgb}{1,1,0}
\definecolor{darkgreen}{rgb}{0,0.6,0}
\definecolor{darkyellow}{rgb}{0.8,0.8,0}
\definecolor{orange}{rgb}{1,0.5,0}
\definecolor{tuc}{RGB}{0,90,70}
\definecolor{tuclight}{RGB}{218,234,194}
\definecolor{tucorange}{RGB}{242,148,0}
\definecolor{tucbg}{RGB}{224,233,233}

\usepackage[
	colorlinks=true,                   
	urlcolor=blue,                     
	filecolor=blue,                    
	linkcolor=blue,                    
	citecolor=blue,                    
	breaklinks=true,                   
	backref=page,                      
	pagebackref=true,                  
	bookmarks=true,                    
	bookmarksopen=false,               
	bookmarksnumbered=true,            
	pdfauthor={Fabian Teichert},       
	pdfdisplaydoctitle=true,           
	pdfstartview=FitH,                 
	pdfpagelayout=TwoPageRight,        
]{hyperref}

\newif\ifsinglepaper\singlepaperfalse
\addto\captionsenglish{
	\renewcommand\bibname{References}
}

\renewcommand*{\backref}[1]{}
\renewcommand*{\backrefalt}[4]{%
	\ifsinglepaper\else%
		\ifcase #1
		\or (cited at p.~#2).
		\else (cited at pp.~#2).
		\fi%
	\fi
}
\usepackage{cite} 

\newcommand{\bstindent}{99}
\newcommand{\bstaddress}{}
\newcommand{\bstauthor}{}

\newcommand{\bstjournal}{}

\newcommand{\bstpublisher}{}

\newcommand{\bsttitle}{\itshape}
\newcommand{\bstvolume}{}
\newcommand{\bstyear}{}
\newcommand{\bbland}{and}

\newcommand\bibliographysection{\section}
\newcommand\bibliographysectionstyle{}
\newcommand\bibliographyitemsize{\normalsize}
\newcommand\bibliographyitemseparation{}
\makeatletter
\newcommand\bibcontentsline{\addcontentsline{toc}{section}{References}}
\renewenvironment{thebibliography}[1]{%
	\bibliographysection{\bibliographysectionstyle\bibname}
	\bibcontentsline%
	\renewcommand\rightmark{\bibname}
	\list{\@biblabel{\@arabic\c@enumiv}}{\settowidth\labelwidth{\@biblabel{#1}}%
		\leftmargin\labelwidth%
		\advance\leftmargin\labelsep%
		\@openbib@code%
		\usecounter{enumiv}%
		\let\p@enumiv\@empty%
		\renewcommand\theenumiv{\@arabic\c@enumiv}%
	}%
	\sloppy
	\clubpenalty4000
	\@clubpenalty \clubpenalty
	\widowpenalty4000%
	\sfcode`\.\@m%
}{%
	\def\@noitemerr{\@latex@warning{Empty `thebibliography' environment}}%
	\endlist%
}
\makeatother
\let\oldthebibliography\thebibliography
\renewcommand\thebibliography[1]{
	\bibliographyitemsize
	\oldthebibliography{#1}
	\bibliographyitemseparation
}

\let\oldtwocolumn\twocolumn
\let\oldonecolumn\onecolumn
\newif\iftwocolumn\twocolumntrue
\def\onecolumn{\twocolumnfalse}
\def\twocolumn{\twocolumntrue}
\newif\ifarticlestyle\articlestylefalse

\renewcommand{\title}[2]{%
	\setcounter{authors}{0}%
	\setcounter{addresses}{0}%
	\setcounter{keywords}{0}%
	\def\inserttitle{#1}%
	\def\articlelabel{#2}
}
\newcommand{\email}[1]{\def\insertemail{#1}}
\newcommand{\abstract}[1]{\def\insertabstract{#1}}
\def\insertjournal{}
\def\insertjournalshort{}
\def\insertdoi{}
\def\insertarxiv{}
\def\insertarxivshort{}
\newcommand{\journal}[4][nothing]{%
	\def\insertjournalshort{#2}%
	\if\relax\detokenize{#3}\relax%
		\def\insertjournal{}%
	\else%
		\def\tmpa{#1}%
		\def\tmpb{submitted}%
		\def\tmpc{accepted}%
		\ifx\tmpa\tmpb%
			\def\journalpre{Submitted to: }%
		\else%
			\ifx\tmpa\tmpc%
				\def\journalpre{Accepted in: }%
			\else%
				\def\journalpre{}%
			\fi%
		\fi%
		\if\relax\detokenize{#4}\relax\def\insertjournal{\journalpre#3}\else\def\insertjournal{\journalpre\href{#4}{#3}}\fi%
	\fi%
}
\newcommand{\doi}[1]{\if\relax\detokenize{#1}\relax\def\insertdoi{}\else\def\insertdoi{DOI: \href{http://dx.doi.org/#1}{#1}}\fi}
\newcommand{\arxiv}[2]{\if\relax\detokenize{#1}\relax\def\insertarxiv{}\def\insertarxivshort{}\else\def\insertarxiv{arXiv: \href{https://arxiv.org/abs/#1}{#1 [#2]}}\def\insertarxivshort{arXiv: #1}\fi}
\newcounter{authors}
\newcounter{addresses}
\newcounter{keywords}
\newcommand{\addauthor}[2]{\csdef{author\arabic{authors}}{#1}\csdef{authoraddress\arabic{authors}}{#2}\stepcounter{authors}}
\newcommand{\addaddress}[1]{\csdef{address\arabic{addresses}}{#1}\stepcounter{addresses}}
\newcommand{\addkeyword}[1]{\csdef{keyword\arabic{keywords}}{#1}\stepcounter{keywords}}

\newcounter{otherchapter}
\newcounter{normalchapter}
\newcounter{othercounter}
\newcounter{i}
\newcounter{j}
\makeatletter
\let\normalchapter\chapter

\renewcommand\chapter{%
	\@ifstar{%
		\normalchapter*%
	}{%
		\stepcounter{normalchapter}%
		\normalchapter%
	}%
}
\newcommand\otherchapter{%
	\protected@write\@auxout{}{\string\@writefile{lof}{\string\addvspace{10\string\p@}}}%
	\protected@write\@auxout{}{\string\@writefile{lot}{\string\addvspace{10\string\p@}}}%
	\scr@startsection{chapter}{1}{\z@}{0ex \@plus -0.2ex}{3.5ex \@plus 0.2ex}{\Large\bfseries}%
}
\newcommand\reftype{}
\newcommand\reflabel{}
\newcommand\refnumber{}
\newcommand\refshortnumber{}
\newcommand\reftext{}
\newcommand{\articletitlesub}{%
	\renewcommand\reftype{chapter}%
	\renewcommand\reflabel{section*.\arabic{othercounter}}%
	\renewcommand\refnumber{\Alph{otherchapter}}%
	\renewcommand\refshortnumber{}%
	\renewcommand\reftext{\inserttitle\ifx\insertjournalshort\empty\ (\insertarxivshort)\else\ (\insertjournalshort)\fi}%
	\pdfbookmark[0]{\reftext}{\reflabel}%
	\otherchapter*{\inserttitle}%
	\protected@write\@auxout{}{\string\@writefile{toc}{\string\contentsline {\reftype}{\string\numberline {\refnumber}\reftext}{\thepage}{\reflabel}}}%
	\Alabel{\articlelabel}%
	\noindent\textbf{%
		\large\csuse{author0}$^{\csuse{authoraddress0}}$%
		\forloop{i}{1}{\value{i} < \value{authors}}{%
			, \csuse{author\arabic{i}}$^{\csuse{authoraddress\arabic{i}}}$%
		}
	}\\[1em]
	\normalsize
	\setcounter{j}{0}
	\forloop{i}{0}{\value{i} < \value{addresses}}{%
		\stepcounter{j}
		$^{\arabic{j}}$\,\csuse{address\arabic{i}}
		\ifthenelse{\value{j}<\value{addresses}}{\\}{}
	}
	\ifx\insertemail\empty\\[1em]\else\\[0.5em]E-mail address: \insertemail\\[1em]\fi
	\textbf{Abstract:} \insertabstract
	\ifthenelse{\value{keywords}=0}{}{
		\\[1em]
		Keywords: \csuse{keyword0}%
			\forloop{i}{1}{\value{i} < \value{keywords}}{%
				; \csuse{keyword\arabic{i}}%
			}
	}
}
\newcommand{\articletitle}{%
	\setcounter{articlepage}{0}%
	\stepcounter{otherchapter}%
	\stepcounter{chapter}%
	\setcounter{section}{0}%
	\setcounter{subsection}{0}%
	\setcounter{subsubsection}{0}%
	\stepcounter{othercounter}%
	\iftwocolumn\oldtwocolumn[\articletitlesub\vspace{1.5em}]\else\oldonecolumn\articletitlesub\fi%
}%
\let\oldchapter\chapter
\let\oldsection\section
\let\oldsubsection\subsection
\let\oldsubsubsection\subsubsection
\newcommand\articlesectiondata[1]{%
	\renewcommand\reftype{section}%
	\renewcommand\reflabel{section.\arabic{chapter}.\arabic{section}}%
	\renewcommand\refnumber{\Alph{otherchapter}.\arabic{section}}%
	\renewcommand\refshortnumber{\arabic{section}}%
	\renewcommand\reftext{#1}%
}
\newcommand\articlesubsectiondata[1]{%
	\renewcommand\reftype{subsection}%
	\renewcommand\reflabel{subsection.\arabic{chapter}.\arabic{section}.\arabic{subsection}}%
	\renewcommand\refnumber{\Alph{otherchapter}.\arabic{section}.\arabic{subsection}}%
	\renewcommand\refshortnumber{\arabic{section}.\arabic{subsection}}%
	\renewcommand\reftext{#1}%
}
\newcommand\articlesectionnostar[1]{%
	\articlesectiondata{#1}%
	\pdfbookmark[1]{\reftext}{\reflabel}%
	\scr@startsection{section}{1}{\z@}{-3.5ex \@plus -1ex \@minus -0.2ex}{2.3ex \@plus 0.2ex}{\normalfont\bfseries}{#1}%
	\protected@write\@auxout{}{\string\@writefile{toc}{\string\contentsline {\reftype}{\string\numberline {\refnumber}#1}{\thepage}{\reflabel}}}%
}
\newcommand\articlesectionstar[1]{%
	\articlesectiondata{#1}%
	\scr@startsection{section}{1}{\z@}{-3.5ex \@plus -1ex \@minus -0.2ex}{2.3ex \@plus 0.2ex}{\normalfont\bfseries}*{#1}%
}
\newcommand\articlesubsectionnostar[1]{%
	\articlesubsectiondata{#1}%
	\pdfbookmark[2]{\reftext}{\reflabel}%
	\scr@startsection{subsection}{2}{\z@}{-3.5ex \@plus -1ex \@minus -0.2ex}{2.3ex \@plus 0.2ex}{\normalfont\bfseries}{#1}%
	\protected@write\@auxout{}{\string\@writefile{toc}{\string\contentsline {\reftype}{\string\numberline {\refnumber}#1}{\thepage}{\reflabel}}}%
}
\newcommand\articlesubsectionstar[1]{%
	\articlesubsectiondata{#1}%
	\scr@startsection{subsection}{2}{\z@}{-3.5ex \@plus -1ex \@minus -0.2ex}{2.3ex \@plus 0.2ex}{\normalfont\bfseries}*{#1}%
}
\newcommand\articlesection{\@ifstar{\stepcounter{othercounter}\articlesectionstar}{\articlesectionnostar}}
\newcommand\articlesubsection{\@ifstar{\stepcounter{othercounter}\articlesubsectionstar}{\articlesubsectionnostar}}
\renewcommand\chapter{\@ifstar{\stepcounter{othercounter}\oldchapter*}{\oldchapter}}
\renewcommand\section{\@ifstar{\stepcounter{othercounter}\oldsection*}{\oldsection}}
\renewcommand\subsection{\@ifstar{\stepcounter{othercounter}\oldsubsection*}{\oldsubsection}}
\renewcommand\subsubsection{\@ifstar{\stepcounter{othercounter}\oldsubsubsection*}{\oldsubsubsection}}
\newcommand\listof{}

\newcommand\articlefiguredata{%
	\renewcommand\listof{lof}%
	\renewcommand\reftype{figure}%
	\renewcommand\reflabel{figure.\arabic{chapter}.\arabic{figure}}%
	\renewcommand\refnumber{\Alph{otherchapter}.\arabic{figure}}%
	\renewcommand\refshortnumber{\arabic{figure}}%
}
\newcommand\articletabledata{%
	\renewcommand\listof{lot}%
	\renewcommand\reftype{table}%
	\renewcommand\reflabel{table.\arabic{chapter}.\arabic{table}}%
	\renewcommand\refnumber{\Alph{otherchapter}.\arabic{table}}%
	\renewcommand\refshortnumber{\arabic{table}}%
}
\renewenvironment{figure}{\articlefiguredata\begin{oldfigure}}{\end{oldfigure}} 
\renewenvironment{table}{\articletabledata\begin{oldtable}}{\end{oldtable}} 
\newenvironment{articlefigure}{\articlefiguredata\begin{figure}}{\end{figure}}
\newenvironment{articlefigure*}{\articlefiguredata\begin{figure*}}{\end{figure*}}
\newenvironment{articletable}{\articletabledata\begin{table}}{\end{table}}
\newenvironment{articletable*}{\articletabledata\begin{table*}}{\end{table*}}
\let\oldcaption\caption
\newcommand\Acaption[2][]{%
	\oldcaption[#1]{#2}%
	\renewcommand\reftext{#1}%
	\protected@write\@auxout{}{\string\@writefile{\listof}{\string\contentsline {\reftype}{\string\numberline {\refnumber}#1}{\thepage}{\reflabel}}}%
}
\renewcommand\caption[2][]{\ifarticlestyle\Acaption[#1]{#2}\else\oldcaption[#1]{#2}\fi}

\newcommand\botholdlabel[1]{\oldlabel{#1}\oldlabel{A#1}}
\newenvironment{articleequation}{\begin{equation}\renewcommand\label{\botholdlabel}}{\end{equation}} 
\newcommand\Aref[1]{\oldref{A#1}}
\newcommand\Alabel[1]{%
	\protected@write\@auxout{}{\string\newlabel{#1}{{\refnumber}{\thepage}{\reftext}{\reflabel}{}}}%
	\protected@write\@auxout{}{\string\newlabel{A#1}{{\refshortnumber}{\thepage}{\reftext}{\reflabel}{}}}%
}
\AtBeginDocument{%
	\let\oldref\ref%
	\let\oldlabel\label%
	\renewcommand\ref[1]{\ifarticlestyle\Aref{#1}\else\oldref{#1}\fi}%
	\renewcommand\label[1]{\ifarticlestyle\Alabel{#1}\else\oldlabel{#1}\fi}%
}
\makeatother

\newcommand\articlestyleheaderleft{%
	\ifx\insertarxiv\empty%
		\ifx\insertjournal\empty\linebreak\textnormal\insertdoi\else\linebreak\textnormal\insertjournal\fi%
	\else%
		\ifx\insertdoi\empty\linebreak\textnormal\insertjournal\else\textnormal\insertjournal\linebreak\textnormal\insertdoi\fi%
	\fi%
}
\newcommand\articlestyleheaderright{%
	\ifx\insertarxiv\empty%
		\ifx\insertjournal\empty\else\linebreak\textnormal\insertdoi\fi%
	\else%
		\linebreak\textnormal\insertarxiv%
	\fi%
}
\newcommand\articlestyleheadercenter{%
	\linebreak\textnormal\thechapter
}

\newcounter{articlepage}
\newcommand\articlepagemark{\arabic{articlepage}}

\newcommand\nocontentsline[3]{}
\let\oldaddcontentsline\addcontentsline
\newcommand\normalstyle{%
	\articlestylefalse%
	\KOMAoptions{fontsize=11pt}%
	\newgeometry{left=3cm,right=2.5cm,top=4cm,bottom=4cm}
	\setlength{\headheight}{26pt}
	\setlength{\headsep}{24pt}
	\setlength{\footskip}{30pt}
	\setlength{\TwoColumnWidth}{\textwidth}
	\setlength{\OneColumnWidth}{0.5\TwoColumnWidth-0.5\columnsep}
	\clearpairofpagestyles%
	\ihead{}%
	\chead{}%
	\ohead{\ifthispageodd{\textnormal\rightmark}{\textnormal\leftmark}}%
	\ifoot{}%
	\cfoot{}%
	\ofoot[\textnormal\pagemark]{\textnormal\pagemark}%
	\let\addcontentsline\oldaddcontentsline%
	\renewcommand{\thechapter}{\arabic{normalchapter}}%
	\renewcommand{\thesection}{\arabic{normalchapter}.\arabic{section}}%
	\renewcommand{\thesubsection}{\arabic{normalchapter}.\arabic{section}.\arabic{subsection}}%
	\renewcommand{\thesubsubsection}{\arabic{normalchapter}.\arabic{section}.\arabic{subsection}.\arabic{subsubsection}}%
	\renewcommand{\thefigure}{\arabic{normalchapter}.\arabic{figure}}%
	\renewcommand{\thetable}{\arabic{normalchapter}.\arabic{table}}%
	\renewcommand{\theequation}{\arabic{normalchapter}.\arabic{equation}}%
	\renewcommand\bibliographysection{\section*}%
	\renewcommand\bibcontentsline{\addcontentsline{toc}{section}{References}}
	\renewcommand\bibliographysectionstyle{}%
	\renewcommand\bibliographyitemsize{\normalsize}%
	\renewcommand\bibliographyitemseparation{%
		\setlength{\parskip}{0pt}%
		\setlength{\itemsep}{5pt plus 0.3ex}%
	}%
	\allowdisplaybreaks%
}
\newcommand\articlestyle{%
	\articlestyletrue%
	\KOMAoptions{fontsize=10pt}%
	\newgeometry{left=1.5cm,right=1.5cm,top=2.95cm,bottom=1.55cm}
	\setlength{\headheight}{24pt}
	\setlength{\headsep}{20pt}
	\setlength{\footskip}{1.2cm}
	\setlength{\TwoColumnWidth}{\textwidth}
	\setlength{\OneColumnWidth}{0.5\TwoColumnWidth-0.5\columnsep}
	\clearpairofpagestyles%
	\ihead{\ifthispageodd{\articlestyleheaderleft}{\articlestyleheaderright}}%
	\chead{\ifsinglepaper\else\articlestyleheadercenter\fi}%
	\ohead{\ifthispageodd{\articlestyleheaderright}{\articlestyleheaderleft}}%
	\ifoot{}%
	\cfoot{\ifsinglepaper\textnormal\pagemark\else\stepcounter{articlepage}\textnormal{\thechapter-\articlepagemark}\fi}%
	\ofoot{\ifsinglepaper\else\textnormal\pagemark\fi}%
	\let\addcontentsline\nocontentsline%
	\renewcommand{\thechapter}{\Alph{otherchapter}}%
	\renewcommand{\thesection}{\arabic{section}}%
	\renewcommand{\thesubsection}{\arabic{section}.\arabic{subsection}}%
	\renewcommand{\thesubsubsection}{\arabic{section}.\arabic{subsection}.\arabic{subsubsection}}%
	\renewcommand{\thefigure}{\arabic{figure}}%
	\renewcommand{\thetable}{\arabic{table}}%
	\renewcommand{\theequation}{\arabic{equation}}%
	\renewcommand\bibliographysection{\articlesection*}%
	\renewcommand\bibcontentsline{\oldaddcontentsline{toc}{section}{References}}
	\renewcommand\bibliographysectionstyle{\normalsize}%
	\renewcommand\bibliographyitemsize{\small}%
	\renewcommand\bibliographyitemseparation{%
		\setlength{\parskip}{0pt}%
		\setlength{\itemsep}{0pt plus 0.3ex}%
	}%
	\interdisplaylinepenalty=10000%
}
\normalstyle

\renewcommand{\hbar}{\mathchar'26\mkern-9mu \mathrm{h}}

\newcommand{\hamilton}{\mathcal{H}}
\newcommand{\overlap}{\mathcal{O}}
\newcommand{\coupling}{\tau}

\newcommand{\green}{\mathcal{G}}

\newcommand{\transmission}{\mathcal{T}}

\newcommand{\imag}{\text{i}}

\newcommand{\eqalign}[1]{\begin{aligned}#1\end{aligned}}

\singlepapertrue
\begin{document}

\raggedbottom

\frontmatter
\clearpairofpagestyles
\ofoot[\textnormal\pagemark]{\textnormal\pagemark}
\KOMAoptions{headsepline=false}

\mainmatter
\KOMAoptions{headsepline=true}

\normalstyle
\articlestyle
\normalsize

\renewcommand{\hamilton}{\mathcal{H}}
\renewcommand{\coupling}{\tau}
\renewcommand{\overlap}{\mathcal{S}}
\renewcommand{\green}{\mathcal{G}}
\renewcommand{\transmission}{\mathcal{T}}
\renewcommand{\imag}{\text{i}}

\twocolumn 

\title{Influence of defect-induced deformations on electron transport in carbon nanotubes}{JPC2}

\addauthor{Fabian Teichert}{1,3,5}
\addauthor{Christian Wagner}{1,2}
\addauthor{Alexander Croy}{4}
\addauthor{J\"org Schuster}{3,5}

\addaddress{Institute of Physics, Chemnitz University of Technology, 09107 Chemnitz, Germany}
\addaddress{Center for Microtechnologies, Chemnitz University of Technology, 09107 Chemnitz, Germany}
\addaddress{Dresden Center for Computational Materials Science (DCMS), TU Dresden, 01062 Dresden, Germany}
\addaddress{Institute for Materials Science and Max Bergmann Center of Biomaterials, Technische Universit\"at Dresden, Dresden, Germany}
\addaddress{Fraunhofer Institute for Electronic Nano Systems (ENAS), 09126 Chemnitz, Germany}

\email{fabian.teichert@physik.tu-chemnitz.de}

\abstract{
We theoretically investigate the influence of defect-induced long-range deformations in carbon nanotubes on their electronic transport properties.
To this end we perform numerical ab-initio calculations using a density-functional-based tight-binding (DFTB) model for various tubes with vacancies.
The geometry optimization leads to a change of the atomic positions.
There is a strong reconstruction of the atoms near the defect (called ``distortion'') and there is an additional long-range deformation.
The impact of both structural features on the conductance is systematically investigated.
We compare short and long CNTs of different kinds with and without long-range deformation.
We find for the very thin (9,0)-CNT that the long-range deformation additionally affects the transmission spectrum and the conductance compared to the short-range lattice distortion.
The conductance of the larger (11,0)- or the (14,0)-CNT is overall less affected implying that the influence of the long-range deformation decreases with increasing tube diameter. 
Furthermore, the effect can be either positive or negative depending on the CNT type and the defect type.
Our results indicate that the long-range deformation must be included in order to reliably describe the electronic structure of defective, small-diameter zigzag tubes.
}

\addkeyword{carbon nanotube (CNT)}
\addkeyword{defect}
\addkeyword{deformation}
\addkeyword{electronic transport}
\addkeyword{density-functional-based tight binding (DFTB)}

\journal{J. Phys. Commun. 2 (2018), 115023}{Journal of Physics Communications 2 (2018), 115023}{http://iopscience.iop.org/article/10.1088/2399-6528/aaf08c} 
\doi{10.1088/2399-6528/aaf08c} 
\arxiv{1705.01753}{cond-mat.mes-hall} 

\articletitle

\articlesection{Introduction}

Carbon nanotubes (CNTs) offer a large variety of properties from metallic types to semiconducting types depending on their structure parameters~\cite{RevModPhys.79.677, NanoRes.1.361, JPhysCondMat.24.313202}.
Because of their well defined quasi one-dimensional structure and their tunable electronic transport properties they are a promising material for future nanoelectronic devices, e.g. semiconducting CNTs may be used as sensor or as transistor channel~\cite{Nature.393.49}.

The fabrication of CNT-based devices leads to inevitable defects.
Better understanding the impact of those defects on the device performance is necessary.
For instance, it is well known that introducing vacancies in a perfect crystal will inhibit the electronic transport.
However, such vacancies will always be accompanied by a defect-induced short-range reconstruction of the lattice, which is named ``distortion'' in the following.
Especially for nano-scale systems, defects may also induce a weak long-range reconstruction which influence the electronic structure~\cite{PhysRevB.65.235412}.
In the following, this is named ``deformation''.

Since the presence of defects (e.g. vacancies, substitutional atoms, functionalizations etc.) has a large impact on the electron transport~\cite{Science.272.523, Nature.382.54, PhysRevB.63.245405, NatureMaterials.4.534, ComputMatSci.93.15, NanoLett.9.2285, JPhysDApplPhys.43.305402, SuperlatticesMicrostructures.98.306}, defects are in the focus of scientific work.
Theoretical studies examine electronic structure and transport properties of defective CNTs, where a geometry optimization is used in order to extract the real structure of CNTs with single defects, periodic defects, or randomly distributed defects~\cite{PhysRevLett.95.266801, JPhysCondMat.20.294214, JPhysCondMat.20.304211, JPhysCondMat.26.045303, JPhysChemC.116.1179, NJPhys.16.123026, SolidStateCommun.149.874, JPhysChemC.117.15266, PhysStatSolB.247.2962, NanoRes.3.288, NanoLett.9.940}.
But within the geometry optimization only short CNTs are typically considered, including the short-range reconstruction of the atoms around the defect.
Usually, it is assumed that this short-range reconstruction constitutes the most significant structural change. Nevertheless, the influence of a long-range deformation on the electron transport has so far been neglected.
Investigations based on continuum elasticity indicate that the range of the vacancy-induced deformation scales with the radius of the tube~\cite{Croy.InPreparation}.

In the following study we investigate the influence of the long-range deformation on the electronic transport properties of carbon nanotubes with vacancies.
We compare the transport through very short/long CNTs to suppress/introduce this long-range deformation.
We use a density-functional-based tight-binding (DFTB) method for the geometry optimization and the electronic structure calculations.
The conductance is computed based on the quantum transport theory described below.

\articlesection{Theory}

\begin{articlefigure*}[t]
	\newcommand\scale{0.075}
	\begin{minipage}{0.07\textwidth}~\end{minipage}%
	\begin{minipage}{0.31\textwidth}\centering Monovacancy (MV)\end{minipage}%
	\begin{minipage}{0.31\textwidth}\centering Divacancy, type 1 (DV1)\end{minipage}%
	\begin{minipage}{0.31\textwidth}\centering Divacancy, type 2 (DV2)\end{minipage}\\[0.5em]
	\begin{minipage}{0.07\textwidth}(9,0)\end{minipage}%
	\begin{minipage}{0.31\textwidth}\includegraphics[scale=\scale]{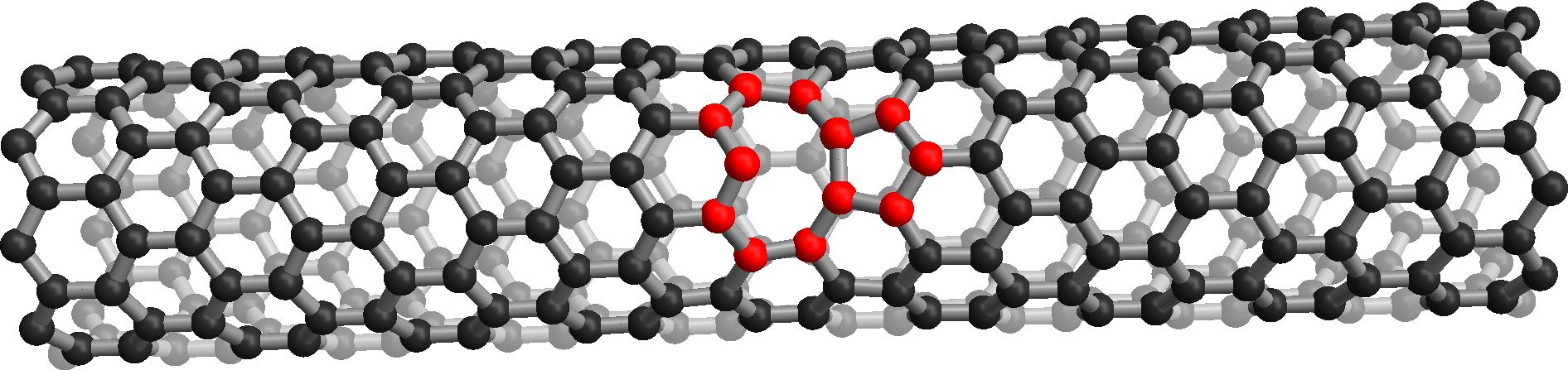}\end{minipage}%
	\begin{minipage}{0.31\textwidth}\includegraphics[scale=\scale]{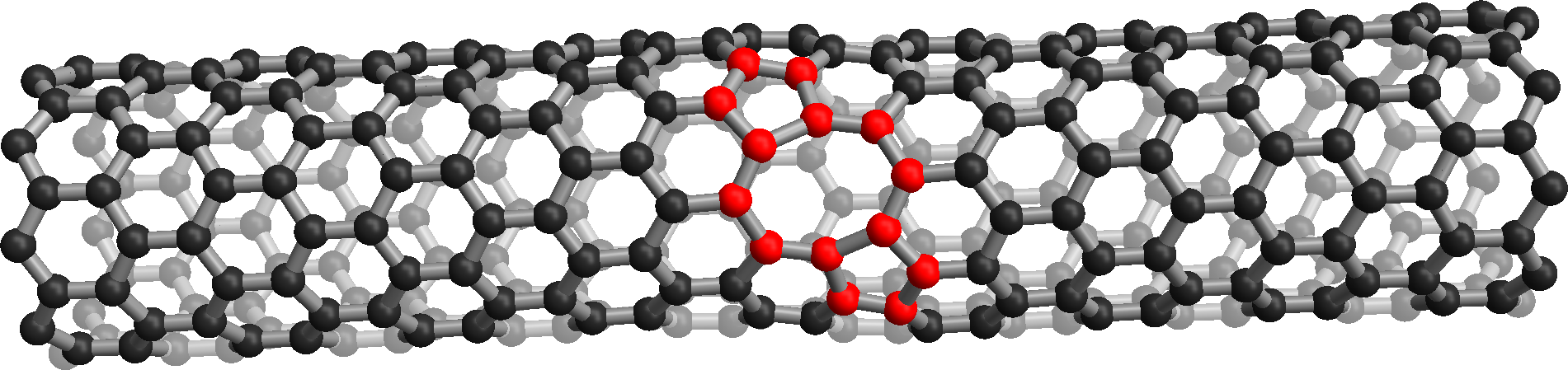}\end{minipage}%
	\begin{minipage}{0.31\textwidth}\includegraphics[scale=\scale]{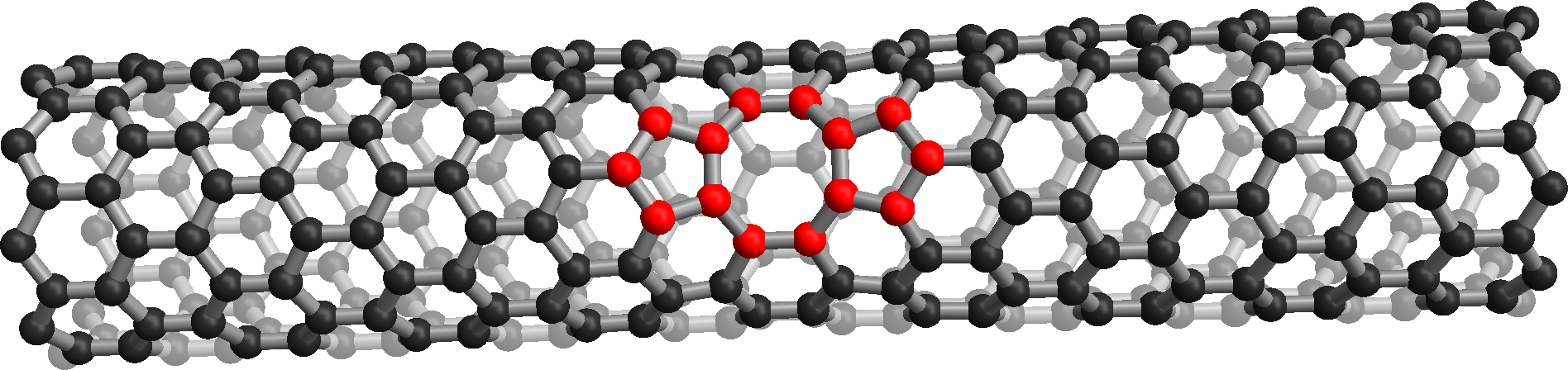}\end{minipage}\\[0.5em]
	\begin{minipage}{0.07\textwidth}(11,0)\end{minipage}%
	\begin{minipage}{0.31\textwidth}\includegraphics[scale=\scale]{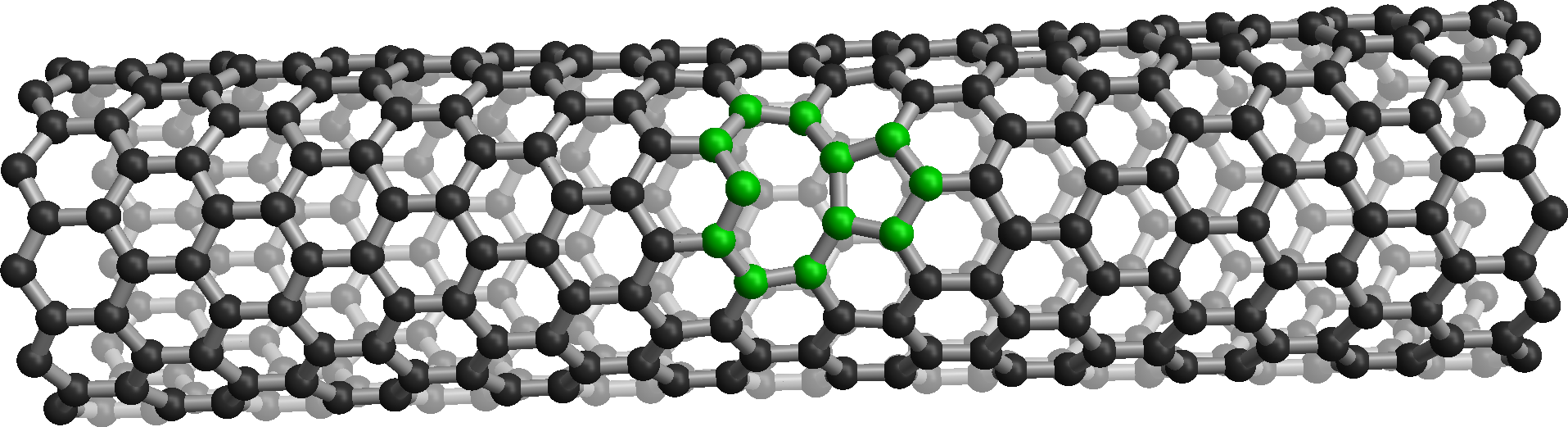}\end{minipage}%
	\begin{minipage}{0.31\textwidth}\includegraphics[scale=\scale]{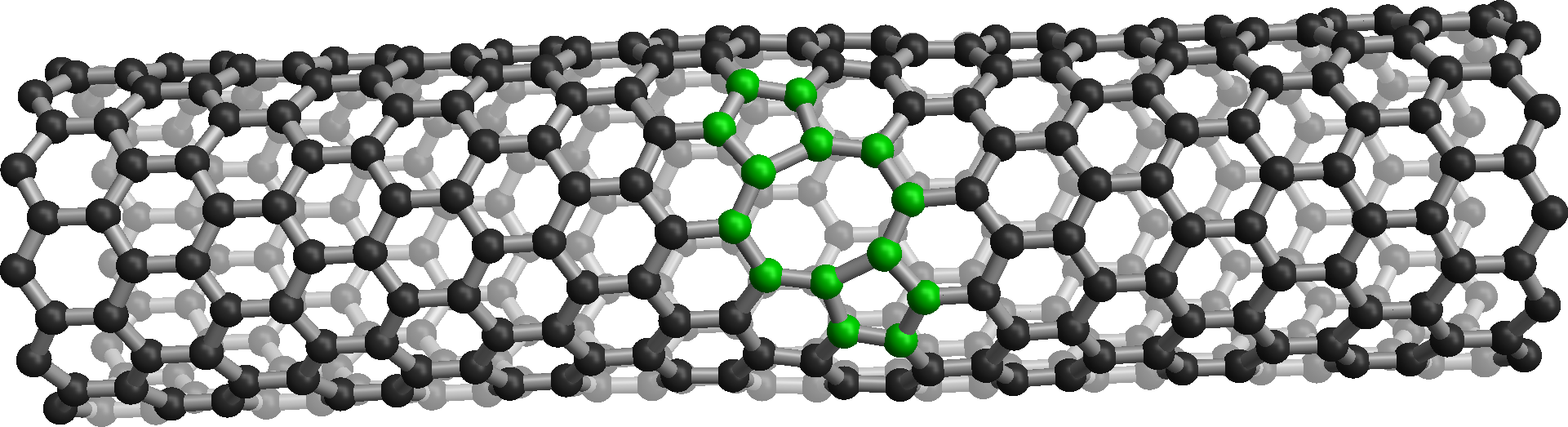}\end{minipage}%
	\begin{minipage}{0.31\textwidth}\includegraphics[scale=\scale]{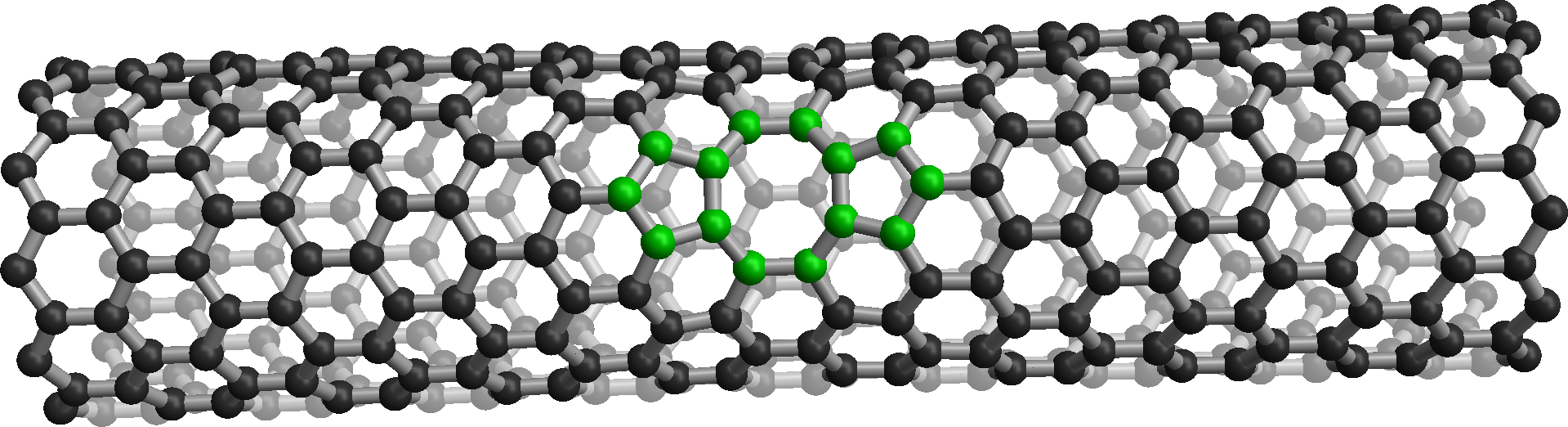}\end{minipage}\\[0.5em]
	\begin{minipage}{0.07\textwidth}(14,0)\end{minipage}%
	\begin{minipage}{0.31\textwidth}\includegraphics[scale=\scale]{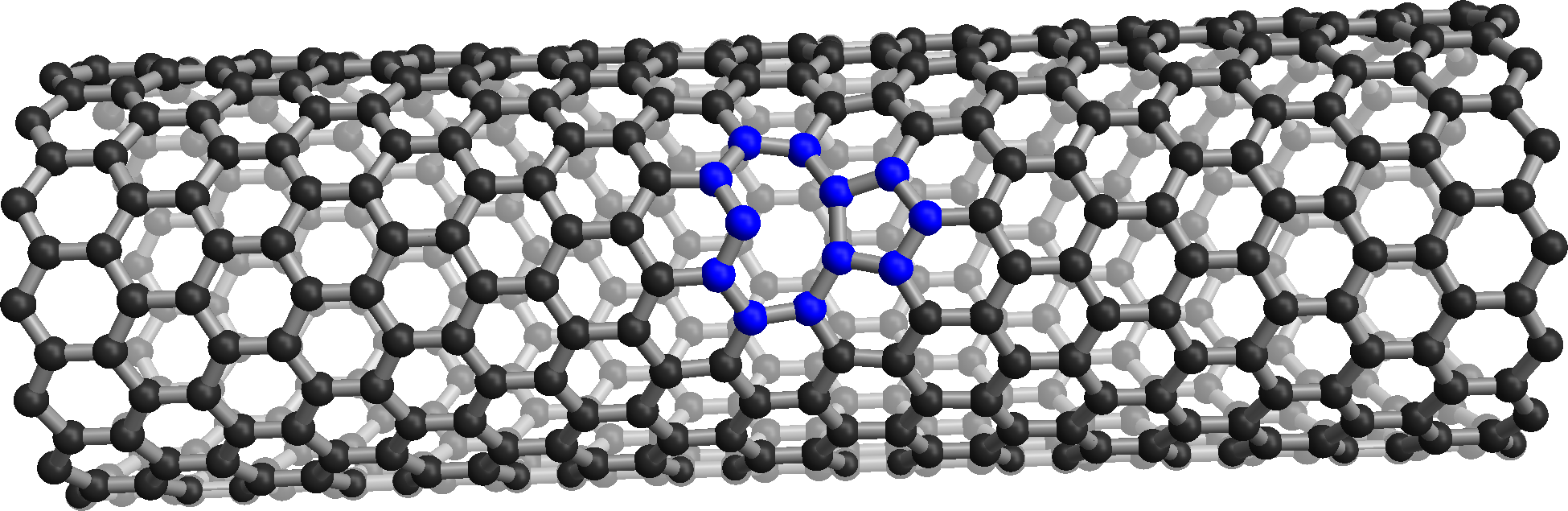}\end{minipage}%
	\begin{minipage}{0.31\textwidth}\includegraphics[scale=\scale]{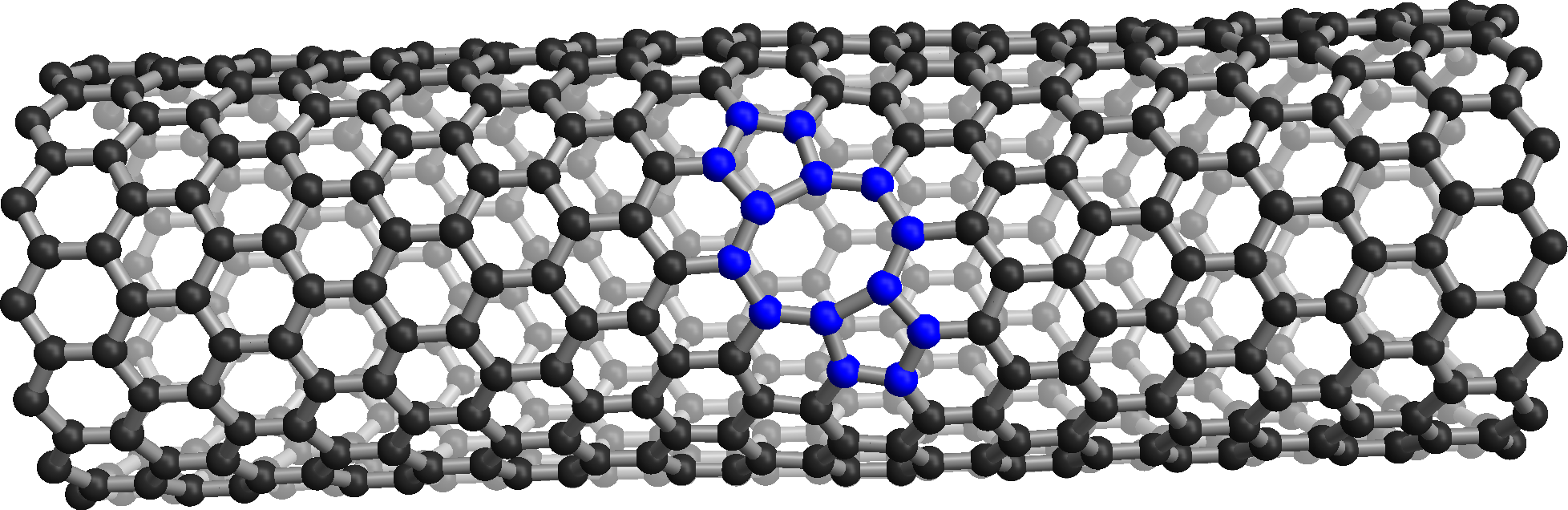}\end{minipage}%
	\begin{minipage}{0.31\textwidth}\includegraphics[scale=\scale]{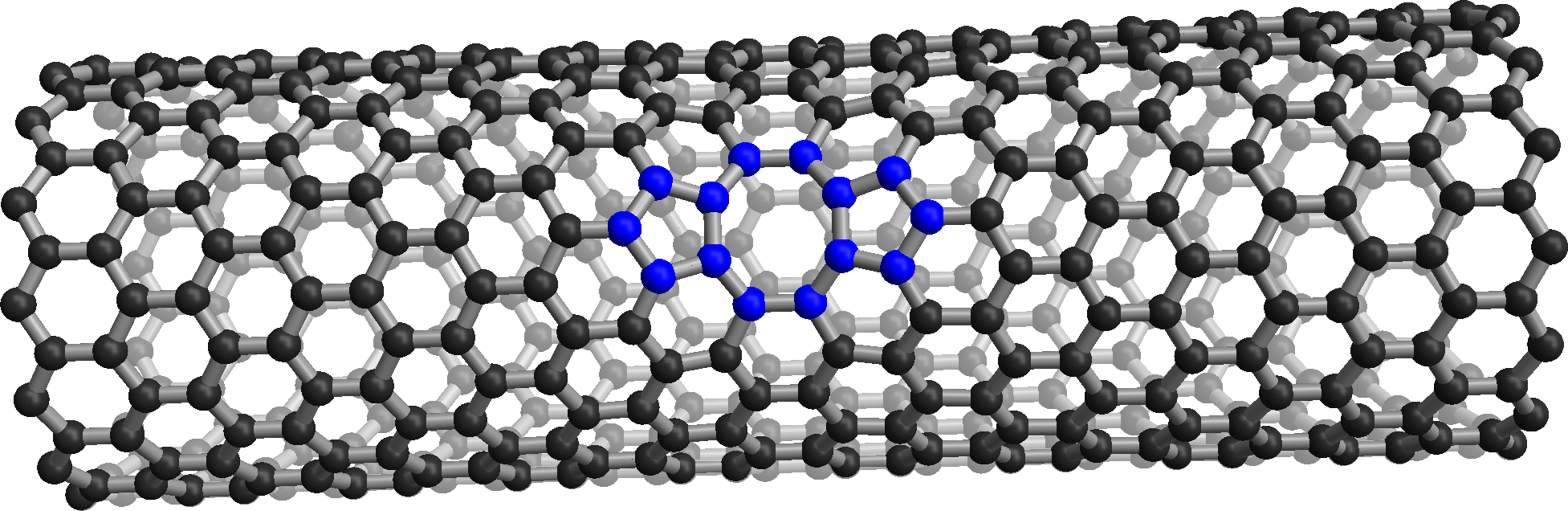}\end{minipage}\\[0.5em]
	\begin{minipage}{0.07\textwidth}(12,4)\end{minipage}%
	\begin{minipage}{0.39\textwidth}\includegraphics[scale=\scale]{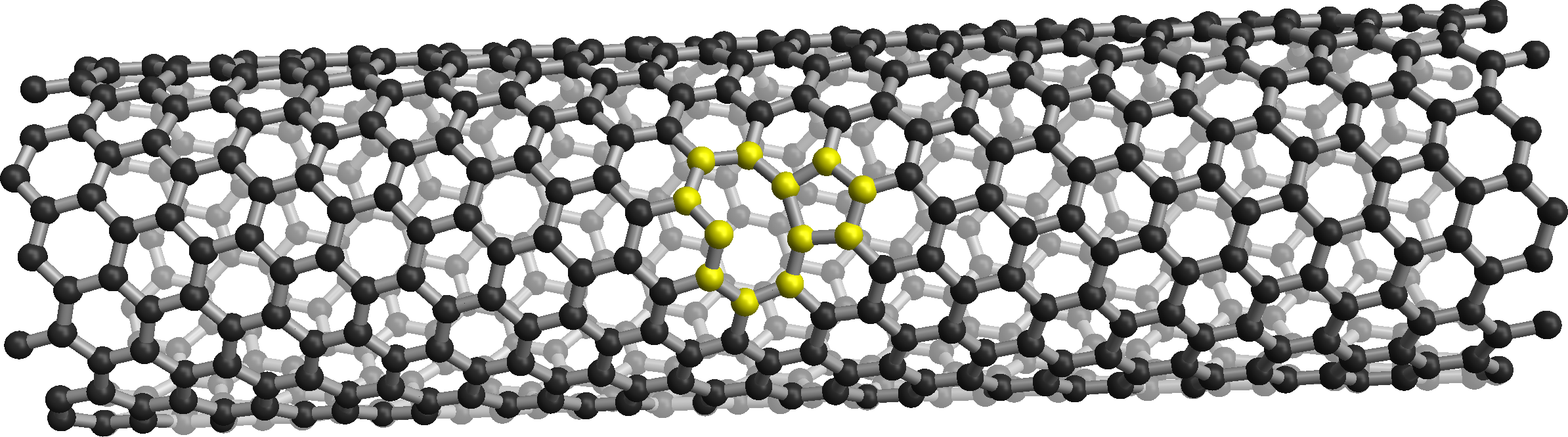}\end{minipage}%
	\begin{minipage}{0.54\textwidth}~\end{minipage}
	\caption[Geometric structures of the CNT defect cells]{(color online) Structures of defective CNTs after the geometry optimization with DFTB for different CNTs and different defects. The pictures show examples with 9 unit cells for the (9,0)-CNT, the (11,0)-CNT, and the (14,0)-CNT and 3 unit cells for the (12,4)-CNT. The highlighted atoms denote the ones around the defect, which are strongly affected by the local reconstruction.}\label{JPC2/fig:cnt}
\end{articlefigure*}

The calculations are performed using a DFTB model \cite{PhysRevB.51.12947, IntJQuantumChem.58.185}.
We use the parameter set 3ob of Gaus et al.~\cite{JChemTheoryComput.9.338}, which includes the 2s and the three 2p Slater-Koster-type orbitals~\cite{PhysRev.94.1498}.
It is tailored for organic molecules with hydrogen, carbon, oxygen, and nitrogen atoms.
It is especially adjusted to benzene, which has the same sp$^2$ structure as the carbon nanotubes we consider.

The electronic part of the DFTB parameter set contains the onsite energies $\epsilon_\mu$ of the orbitals $\mu\in\{s,p_\text{x},p_\text{y},p_\text{z}\}$, the distance-dependent hopping energies $\tau_{\mu\nu}(r)$ between two atomic orbitals ($\mu$, $\nu$) at distance $r$, the one-center-integrals $\sigma_\mu$ of the orbital overlap, and the respective two-center-integrals $\sigma_{\mu\nu}(r)$.
The DFTB Hamiltonian matrix $\hamilton$ and overlap matrix $\overlap$ is given by
\begin{eqnarray}
	\eqalign{
		\hamilton &= \sum_{i\mu}\left| i\mu\right\rangle\epsilon_\mu\left\langle i\mu\right| + {\sum_{i\mu,j\nu}}'\left| i\mu\right\rangle\tau_{\mu\nu}(r_{ij})\left\langle j\nu\right| \quad,\\
		\overlap &= \sum_{i\mu}\left| i\mu\right\rangle\sigma_\mu\left\langle i\mu\right| + {\sum_{i\mu,j\nu}}'\left| i\mu\right\rangle\sigma_{\mu\nu}(r_{ij})\left\langle j\nu\right| \quad,
	}
\end{eqnarray}
where $i$ and $j$ are the atomic indices.
The nuclei part of the DFTB parameter set contains the repulsive potential $U(r)$ of two different nuclei at distance $r$.

The atomic configuration is obtained by a geometry optimization through an energy minimization.
To this end, the corresponding electronic force and the nuclei force at atom $j$ are calculated by
\begin{eqnarray}
	\eqalign{
		\vec{F\,}\!_j^\text{el} &= \sum_{k\in\text{occ.}} \left\langle k\right| \varepsilon_k\nabla\!_j\overlap - \nabla\!_j\hamilton \left| k\right\rangle \quad ,\\
		\vec{F\,}\!_j^\text{nuc} &= -\sum_{k\neq j} \nabla\!_j U(r_{jk}) \quad .
	}
\end{eqnarray}

The electronic transport is described by the transmission spectrum, which is calculated using quantum transport theory~\cite{Datta2005}.
For this, a quasi one-dimensional device configuration is considered where a finite central scattering region is connected to two semi-infinite electrodes.
In our case, the central scattering region contains the CNT with the defect and the electrodes are the ideal CNT.
The Green's function of the central region is
\begin{articleequation}
	\green = ( E\overlap - \hamilton - \varSigma_\text{L} - \varSigma_\text{R} )^{-1}
\end{articleequation}%
with matrices $\varSigma_\text{L/R}$ describing the energy shift of the electronic states due to the electrode coupling.
They are calculated iteratively via the renormalization decimation algorithm~\cite{JPhysFMetPhys.14.1205, JPhysFMetPhys.15.851}.
By using $\green$, the transmission spectrum can be calculated via
\begin{articleequation}
	\mathcal{T}(E) = \text{Tr}(\varGamma_\text{R}\green\varGamma_\text{L}\green^\dagger)
\end{articleequation}%
with matrices $\varGamma_\text{L/R}$ describing the broadening of the electronic states due to the electrode coupling.
Finally, the conductance can be calculated with the Landauer-B\"uttiker formula~\cite{PhysRevB.31.6207}
\begin{articleequation}
	G = -\frac{2\text{e}^2}{\text{h}} \int\limits_{-\infty}^\infty \mathcal{T}(E)f'(E)\,\text{d}E \quad ,
\end{articleequation}%
where $f'(E)$ is the derivative of the Fermi distribution.

\articlesection{Results}

In the following we focus on the long-range deformation in CNTs caused by defects.
The geometric structures which we obtained after the geometry optimization are shown in figure \ref{JPC2/fig:cnt}.
We consider the following different CNT types: (9,0), (11,0), (14,0), and (12,4).
This includes semi-metallic CNTs as well as semiconducting CNTs and zigzag CNTs as well as chiral CNTs, offering at least one tube per type.
For the defects, we focus on the following vacancies: monovacancies (MV), divancancies which are aligned diagonally (DV1), and divacancies which are aligned in parallel (DV2) to the tube axes.
Furthermore, we consider four different CNT lengths:\linebreak $\ell=12.79$\,\AA, 38.36\,\AA, 63.94\,\AA, 89.51\,\AA\ for the ($m$,0)-CNTs resp. $\ell=15.37$\,\AA, 46.11\,\AA, 76.84\,\AA, 107.58\,\AA\ for the (12,4)-CNT.
We label them S/M/L/XL-CNT, according to table \ref{JPC2/tab:length}.
Figure \ref{JPC2/fig:cnt} depicts the M-CNTs.

\begin{articlefigure*}[t]
	\includegraphics{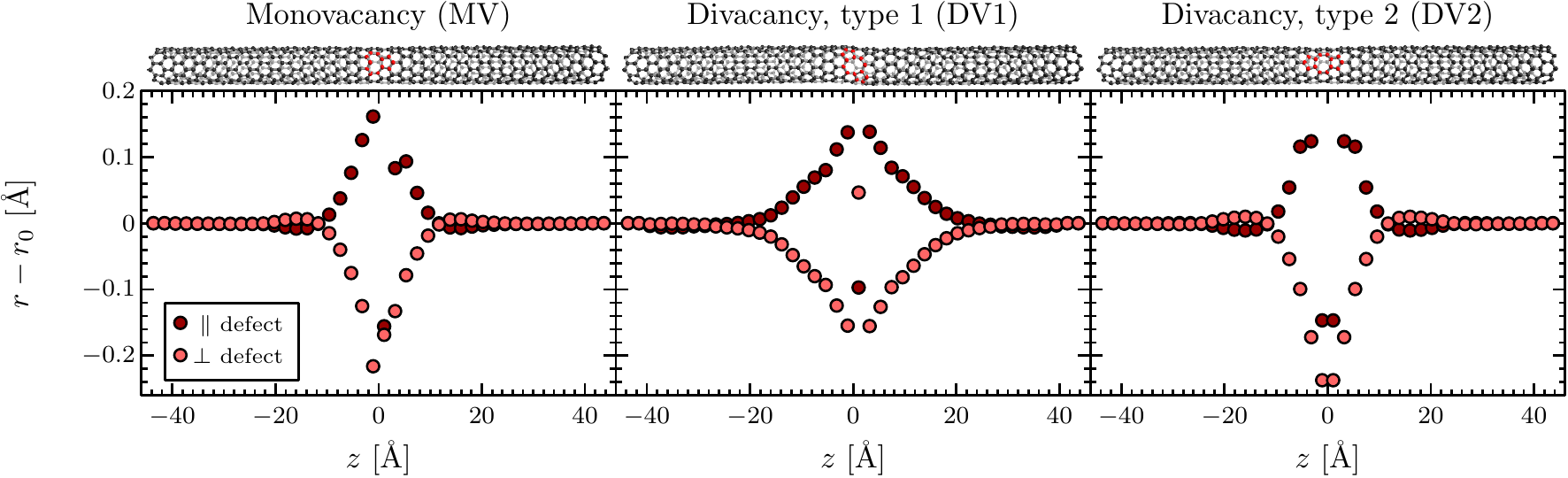}
	\caption[Ellipse half-axes as a function of the axial coordinate]{(color online) Half-axes of the elliptical regression of each carbon ring as a function of the axial coordinate for the XL-(9,0)-CNT for different defects. The whole defective CNT with the correct length scale is shown above.}\label{JPC2/fig:radii}
\end{articlefigure*}

\begin{articletable}[t]
	\begin{tabular}{c|rr|rr}
		\multirow{2}{*}{Notation} & \multicolumn{2}{c|}{($m$,0)-CNT}& \multicolumn{2}{c}{(12,4)-CNT} \\
		& {\#}UC & length & {\#}UC & length \\
		\hline
		S  &  3 & 12.79\,\AA & 1 &  15.37\,\AA \\
		M  &  9 & 38.36\,\AA & 3 &  46.11\,\AA \\
		L  & 15 & 63.94\,\AA & 5 &  76.84\,\AA \\
		XL & 21 & 89.51\,\AA & 7 & 107.58\,\AA
	\end{tabular}
	\caption[Notation of the different CNTs]{Notation of the different CNTs, their length and their corresonding number of unit cells (UC).}\label{JPC2/tab:length}
\end{articletable}

The strong short-range distortion due to the presence of the defect introduces a local reconstruction of the atoms surrounding the defect.
This leads to a deformation influencing the long-range structure.
The MV causes a perturbation of the sp$^2$ hybridization leading to an atom moving outwards the tube and a dented region around the nonagon-pentagon border.
The DV1 introduces a kink, which increases with decreasing tube diameter.
The octagon ring of the DV2 causes a dented region.
These local perturbations cause a global long-range deformation:
The atoms are not displaced as drastic as during the local reconstruction, but the additional change is sufficiently large enough to influence the electronic properties.
The long-range deformation has an ellipse-like shape with a larger ellipticity near the defect and a smaller ellipticity far away from the defect.
In order to get a quantitative idea about the effect, elliptic regressions have been performed for small pieces of the CNT to extract the half-axes as a function of the axial position.
The results are visualized in figure \ref{JPC2/fig:radii} for the (9,0)-CNT, proving the statement that the ellipticity decreases with increasing distance from the defect.
The behavior of other CNTs is qualitatively identical, but quantitatively depends on the CNT type as well as the defect type.
A comprehensive study with more detailed insights how to describe the structure can be found in~\cite{Croy.InPreparation}.

In the following we focus on the electronic transport properties.
In order to quantify the influence of the previously discussed long-range deformation, we calculate the the transmission spectra of the CNTs of figure \ref{JPC2/fig:cnt} for the four different CNT lengths shown in table \ref{JPC2/tab:length}.
We compare the results concerning the tube length, where three different deformation patterns are observed:
(A) The S-CNTs only contain the strong short-range reconstruction around the defect.
The global long-range deformation of the CNT is suppressed.
(B) The M-CNTs additionally contain the decreasing ellipticity, which is the strongest part of the global long-range deformation.
(C) The L-CNTs and the XL-CNTs additionally contain the overshoots of the global long-range deformation shown in figure \ref{JPC2/fig:radii}.
As the XL-CNTs contain all the deformations, they are the reference systems for discussing the influence of neglecting the deformations within the S/M/L-CNTs.

\begin{articlefigure*}[t]
	\includegraphics{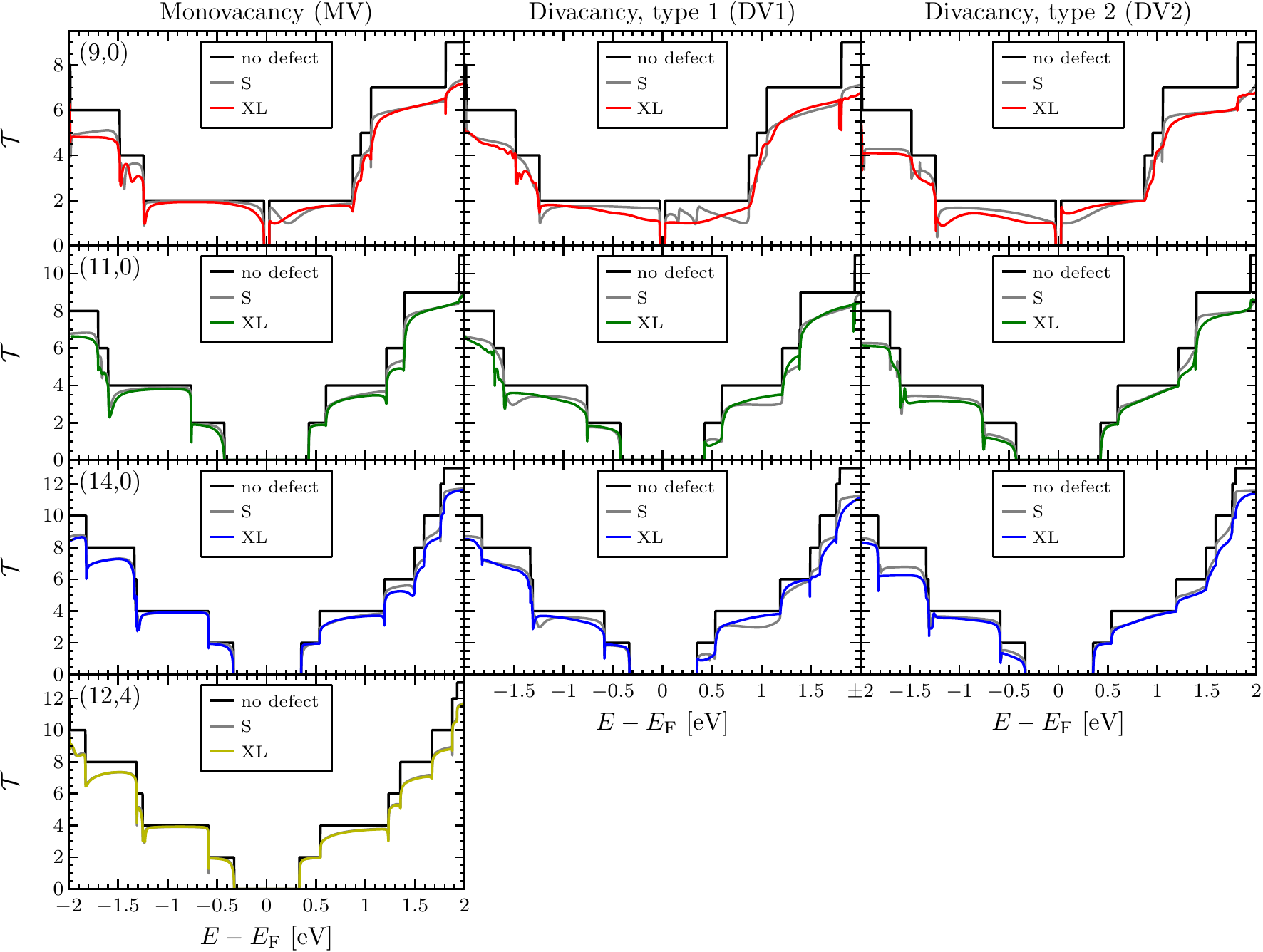}
	\caption[Transmission spectrum for different CNTs with different defects]{(color online) Transmission spectrum for different CNTs with different defects. The color denotes the CNT type according to figure \ref{JPC2/fig:cnt}. Each subgraph shows the transmission spectrum of the ideal CNT (black), the defective S-CNT with only the short-range reconstruction (gray), and the defective XL-CNT with the long-range deformation (color).}\label{JPC2/fig:transmission}
\end{articlefigure*}

\begin{articlefigure*}[t]
	\includegraphics{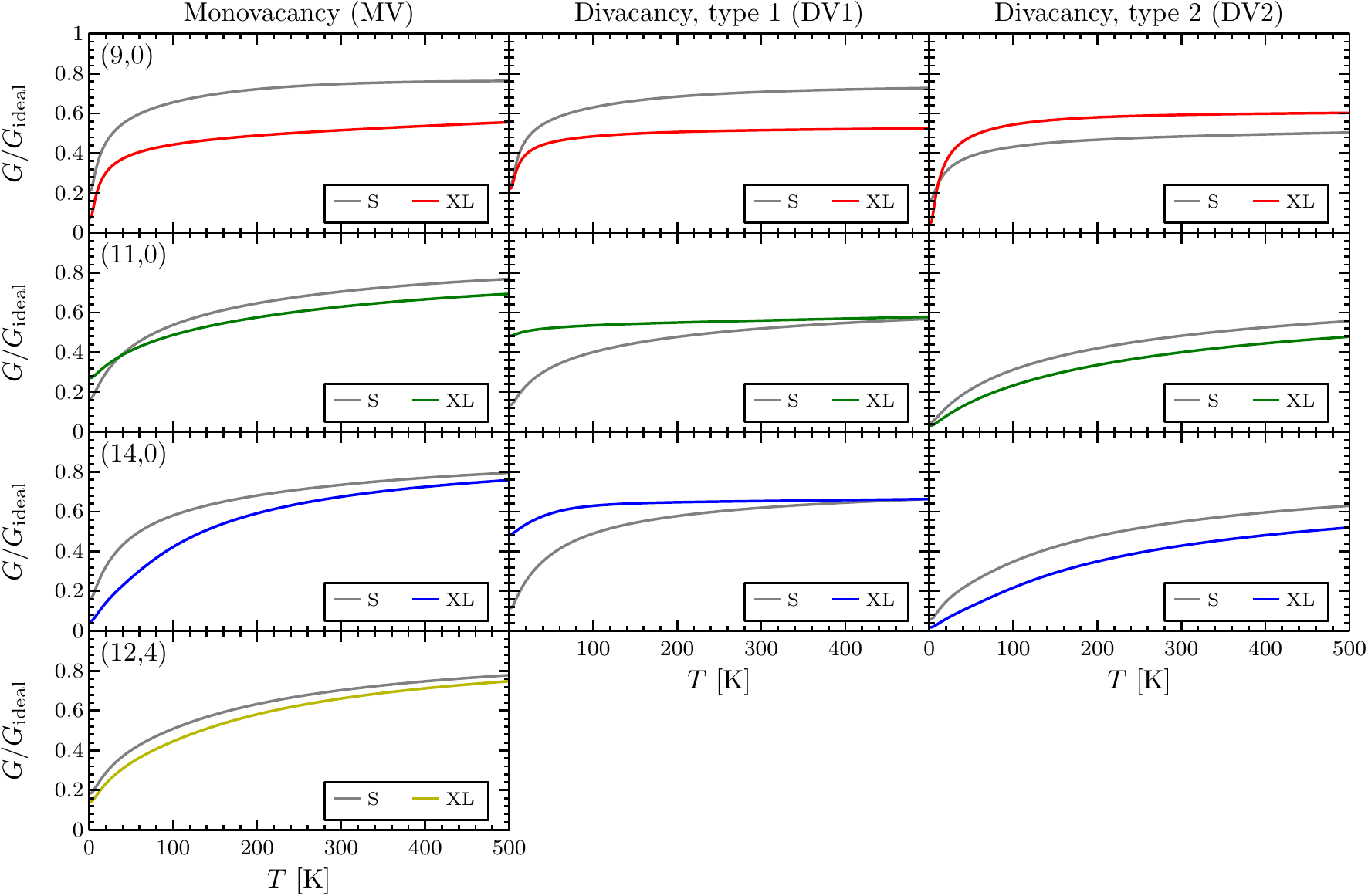}
	\caption[Relative conductance for different CNTs with different defects]{(color online) Relative conductance of the defective CNT normalized to the conductance of the ideal CNT as a funtion of temperature for different CNTs with different defects. The color denotes the CNT type according to figure \ref{JPC2/fig:cnt}. Each subgraph shows the relative conductance of the defective S-CNT with only the short-range reconstruction (gray) and the defective XL-CNT with the long-range deformation (color). The value 1 is the upper limit and corresponds to the ideal case.}\label{JPC2/fig:conductance}
\end{articlefigure*}

The transmission spectra for the S-CNTs and the\linebreak XL-CNTs are depicted in figure \ref{JPC2/fig:transmission}.
For the defective\linebreak S-CNTs the transmission is reduced with respect to the ideal CNT due to the presence of the defect.
Pronounced large dips are present at the band edges.
The comparison of the XL-(9,0)-CNTs with its short counterpart (S) shows the following:
The transmission changes for all three defect types.
Mostly, the transmission is lowered, which is expected due to the additional perturbation of the\linebreak sp$^2$ lattice.
But there are a few energy regions, where the transmission increases, i.e. around $0.3$\,eV for the MV.
The reason could be a detuning of the electronic resonances on a larger length scale.
Nevertheless, this effect is minor and there is no systematic access for its description.
The states near the conduction and valence band edge have the largest impact on the conductance.
The (9,0)-CNT with a MV or a DV1 defect shows a reduced transmission resulting in lower conductance for the CNT with long-range deformation.
The transmission for the (9,0)-CNT with a DV2 defect is slightly lower at the valence band and much higher at the conduction band, leading to an overall increase of the conductance.
Qualitatively similar results can also be observed for the (11,0)- and the (14,0)-CNT.
In most energy regions, the transmission is lower, but not everywhere.
Also here, the change of the conductance can be positive or negative, depending on the transmission near the band edges.
For the (11,0)-MV and the (14,0)-MV the transmission is slightly lowered at the valence band and slightly raised at the conduction band.
For the (11,0)-DV1 and the (14,0)-DV1 it is the other way round.
The diagram in figure \ref{JPC2/fig:transmission} does not reveal whether the changes of the transmission at the valence band or at the conduction band dominate the influence on the conductance.
The result for the (11,0)-DV2 and the (14,0)-DV2 are clearer.
Here, the transmission is lowered at both bands, leading to a lowered conductance.
In addition to the ($m$,0)-CNTs, the (12,4)-CNT with a MV defect as a representative of the defective chiral tubes is shown in figure \ref{JPC2/fig:transmission}.
For this CNT, only very slight and hardly visible differences between the transmission spectra of the S-CNT and the XL-CNT are present, leading to a negligible change in the conductance.
This is due to the fact, that the MV defect does not introduce a long-range deformation into the (12,4)-CNT within our DFTB calculations.
A reason could be that the intrinsic strain along the chiral directions can be better reduced by the local reconstruction than along the circumference direction of the zigzag tubes.
But the calculations are not sensitive enough to show whether this is true or not:
The effective forces are too small to discriminate the absence of the deformation effect from unconverged DFTB geometries.
Finally, the transmission spectra of the M-tubes and the L-tubes are nearly the same as the ones for the XL-tubes (and thus not shown in figure \ref{JPC2/fig:transmission}).
That means that the electron transport is mainly influenced by the CNT part with the exponentially decreasing ellipticity of the global long-range deformation.
Smaller structural features beyond this are not relevant for the electronic structure and transport properties of the CNTs.

To get a more quantitative view on how the conductance is influenced, the ratio of the defective and the ideal case is depicted in figure \ref{JPC2/fig:conductance}.
The previous statements made for the transmission spectra are clearer.
For the MV defect as well as for the DV1 defect the quantitative comparison of the three CNT types (9,0), (11,0), and (14,0) shows a decreasing influence of the long-range deformation on the transmission with increasing diameter.
This could be explained by the fact that the absolute change of the radii is similar for these CNTs.
But the relative values, which are decreasing with increasing diameter, lead to a smaller local distortion of the structure, i.e. a smaller ellipse eccentricity.
In contrast, a decreasing influence of the long-range deformation on the transmission with increasing diameter can not be seen for the DV2 defect.

\begin{articlefigure}[t]
	\includegraphics{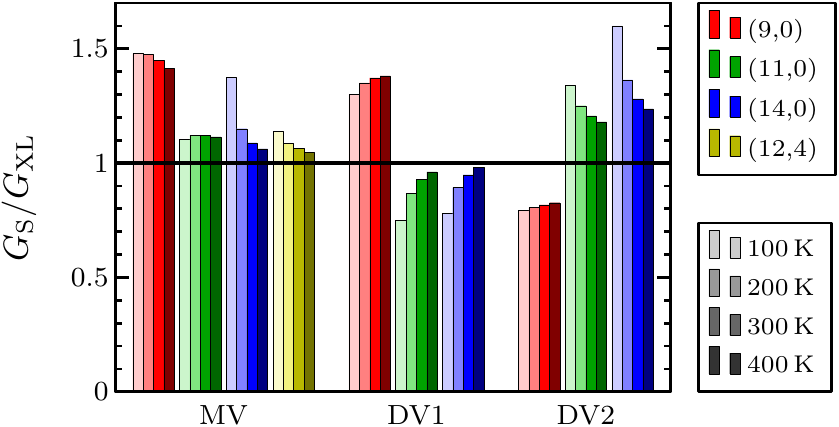}
	\caption[Conductance ratio of CNTs with and without long-range deformation]{(color online) Conductance ratio of the S-CNTs (without the long-range deformation) and the XL-CNTs (with the long-range deformation) for different defects. The color denotes the CNT type according to figure \ref{JPC2/fig:cnt}. Different brightness levels correspond to different temperatures.}\label{JPC2/fig:conductance:relative:T}
\end{articlefigure}

\begin{articletable}[t]
	\begin{tabular}{r|r|r|r}
		~ & MV & DV1 & DV2 \\
		\hline
		(9,0)  & $45\%$ & $37\%$ & $-18\%$ \\
		(11,0) & $12\%$ & $-7\%$ & $21\%$  \\
		(14,0) & $9\%$  & $-5\%$ & $28\%$  \\
		(12,4) & $6\%$  & -------- & --------
	\end{tabular}
	\caption[Relative conductance deviations for neglected long-range deformations]{Relative deviations $G_\text{S}/G_\text{XL}-1$ of the conductance of the S-CNTs (without the long-range deformation) compared to the XL-CNTs (with the long-range deformation) at room temperature (300\,K).}\label{JPC2/tab:conductance:relative}
\end{articletable}

All the curves of figure \ref{JPC2/fig:conductance} are summarized in figure~\ref{JPC2/fig:conductance:relative:T}.
Here, the conductance ratio of the S-CNTs (without the long-range deformation) and the XL-CNTs (with the long-range deformation) is shown.
The resulting relative deviations $G_\text{S}/G_\text{XL}-1$ of the conductance at room temperature (300\,K) are also listed in table \ref{JPC2/tab:conductance:relative}.
The following statements can be extracted:
The conductance of the S-CNT is larger than the XL-CNT for the (9,0)-MV, the (9,0)-DV1, the (11,0)-MV, the (11,0)-DV2, the (14,0)-MV, the (14,0)-DV2, and the (12,4)-MV.
In contrast, the conductance of the S-CNT is smaller than the XL-CNT for the (9,0)-DV2, the (11,0)-DV1, and the (14,0)-DV1.
At room temperature, the effect is significantly large for the (9,0)-MV, the (9,0)-DV1, the (9,0)-DV2, the (11,0)-DV2, and the (14,0)-DV2.
The effect is small for the other ones.
This finally means that the effect is strong for small-diameter CNTs and/or CNTs with the DV2 defect -- the divacancy that points towards the transport direction.
There is no systematic trend concerning the sign of the change.
The conductance of the S-CNT can be larger than the one of the XL-CNT as the additional long-range deformation introduces additional electron scattering.
But it can also be the other way around if the long-range deformation enables a further short-range reconstruction which can also lower the electron scattering.

\begin{articlefigure}[t]
	\includegraphics{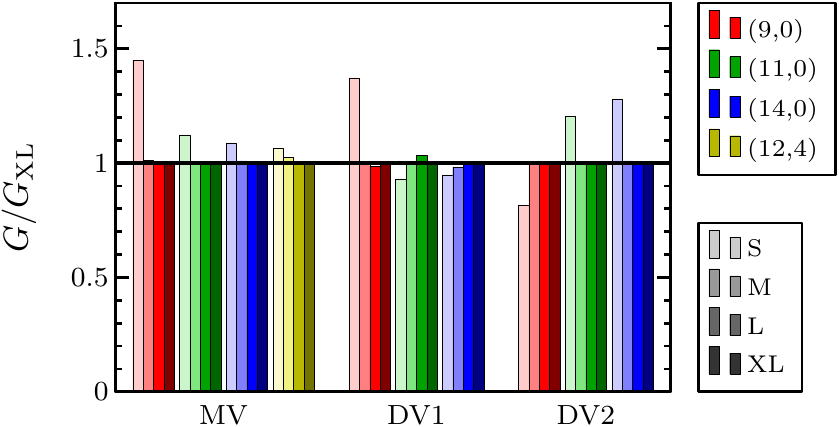}
	\caption[Conductance of CNTs with different lengths]{(color online) Conductance ratio of the S/M/L/XL-sized CNTs and the XL-CNTs (with the long-range deformation) for different defects at room temperature (300\,K). The color denotes the CNT type according to figure \ref{JPC2/fig:cnt}. Different brightness levels correspond to different CNT lengths.}\label{JPC2/fig:conductance:relative:L}
\end{articlefigure}

Figure \ref{JPC2/fig:conductance:relative:L} shows the conductance at 300\,K for all CNT lengths (S, M, L, XL), normalized to the one of the XL-CNT.
The change between the S- and the M-CNT is the largest.
This indicates that the structural feature of the M-sized unit cell, which is the exponential decaying ellipticity, is the most relevant for the conductance.
Further structural changes in the L-CNT and the XL-CNT are the overshoots of the ellipse radii for large distances from the defect as depicted in figure \ref{JPC2/fig:cnt}.
As explained before, the influence of these small features on the transmission spectrum is very small, and thus, their influence on the conductance is negligible.

\articlesection{Summary and conclusions}

We investigated the influence of the defect-induced long-range deformation on the electronic transport properties of CNTs compared to the general impact of the short-range lattice distortion.
For this, we calculated the transmission spectrum and the conductance for CNTs with three different defect types.
For each defect, two geometry optimizations were carried out: One geometry optimization was performed for a short tube in order to suppress the long-range deformation and another one for a very long tube resulting in an elliptic long-range deformation.
It was found that the impact of the long-range deformation on the conductance of small-diameter CNTs is more significant than for CNTs with larger diameter.
This can be explained by a stronger relative change of the CNT curvature due to the presence of the long-range deformation.
Nevertheless, general quantitative statements for different CNT types and defect types are hardly possible.
The effect can either be negative or positive or even not present depending on the considered CNT structure.
Furthermore, the geometry optimization of the medium-long CNTs, which includes the local rearrangement and the main features of the long-range deformation, is sufficient to describe the impact on the transport properties of CNTs.
The influence of smaller features beyond this does not play a role.
This obviously implies the existence of a characteristic defect-dependent length scale.
This length scale is not only characteristic for the lattice distortions of a CNT, but also has an impact on its electronic structure in the deformed region.

We finally conclude that the long-range deformation caused by the presence of defects has to be taken into account for theoretical calculations to get better quantitative statements, especially for small-diameter CNTs.
Most theoretical investigations about the influence of defects on whatever property only deal with the strong short-range lattice distortion, which is indeed the most significant one.
The long-range deformation is usually neglected, but this study shows that it has a significant impact for some defect systems.
These results enable a more realistic picture of transport through defective CNTs, e.g. for studies of application-oriented device simulations.

\articlesection*{Acknowledgment}

This work is funded by the European Union (ERDF) and the Free State of Saxony via the ESF project\linebreak 100231947 (Young Investigators Group Computer Simulations for Materials Design - CoSiMa).
This work is further supported by the German Research Association (DFG) within the Research Unit 1713.
\begin{center}
	\includegraphics[width=0.35\textwidth]{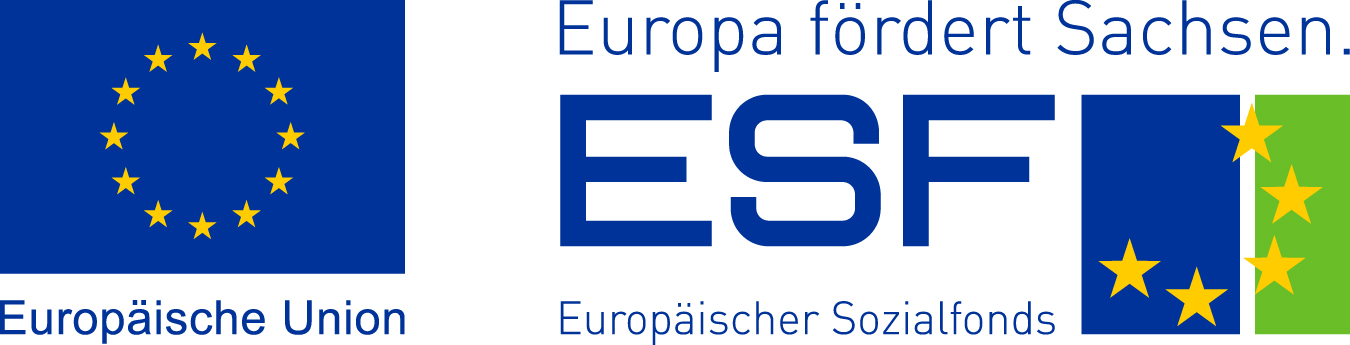}
\end{center}
The publication costs of this article were funded by the German Research Foundation/DFG and the Technische Universit\"at Chemnitz in the funding programme Open Access Publishing.

\end{document}